\documentclass[12pt]{article}
\usepackage{amsmath}
\usepackage{graphicx}
\title{A low cost technique to find gravitational acceleration}

\author{Chetan Kotabage\\
Center of Astrophysics and Department of Physics\\
KLS Gogte Institute of Technology\\
Belagavi, Karnataka, India .590008\\
Email: cvkotabage@git.edu}

\begin{document}
\maketitle

\begin{center}
\textbf{Abstract}\\
\end{center}
The most common way to find gravitational acceleration, g, in a laboratory is to use a simple pendulum and a clock. Alternately, g can be calculated by measuring time and distance for a free fall. Since the time of free fall in a laboratory is short, there are challenges in measuring the short period. In this article, a low cost technique to measure the time for a free fall using a discharging RC circuit is discussed.

\textbf{keywords:} gravitational acceleration, discharging RC circuit

\section{Introduction}
In physics laboratory, acceleration due to gravity can be measured by measuring time period for oscillations of a simple pendulum for a small angle displacement. Alternately, a straight forward way  of measuring acceleration due to gravity is to measure time and distance of a free fall.  For a free fall of one meter, the time of fall is 0.452 s, which is impossible to measure by a usual clock. Nevertheless, commercial experimental set up with a software is available to measure such short period. In a low-cost technique, a discharging RC circuit is used to measure the time of free fall. Such technique not only makes student to understand working of RC circuit but also apply it to find gravitational acceleration. The experimental set up discussed in this article is different from reference \cite{1}, which uses a charging RC circuit.

\section{ Theory of the experiment to find g}
In a free fall motion, an object is released from rest and it covers a distance $d$ in time $t$ before it hits the ground. The kinematic equation of motion for such free fall yields acceleration due to gravity  \cite{2}
\begin{equation}
g=\frac{2d}{t^2}\;.
\label{g}
\end{equation}
For a discharging RC circuit, the voltage of the capacitor varies as \cite{2}
\begin{equation}
V=V_0e^{-\frac{t}{RC}}\;,
\end{equation}
where $V_0$ is the initial voltage of the charged capacitor, $R$ is the resistance and $C$ is the capacitance of the resistor and capacitor respectively.   Thus, the time measured for decrease in the voltage of the capacitor is
\begin{equation}
t=RC\ln \Bigg(\frac{V_0}{V}\Bigg)\;.
\label{RC}
\end{equation}
The low-cost technique involves measurement of time of fall for a ball using a discharging RC circuit.
\section{The experimental set up}
The  circuit diagram of the experimental set up is shown in the figure 1. The charging of the capacitor is done by the turning on the switch $X$. At this point the switch $Y$ is off while the switch $Z$ is on. Once the capacitor gets fully charged, the switch $X$ is tuned off and voltage $V_0$ is recorded. 

\begin{figure}[h]
\centering
\includegraphics[width=6.16cm, height=5.59cm]{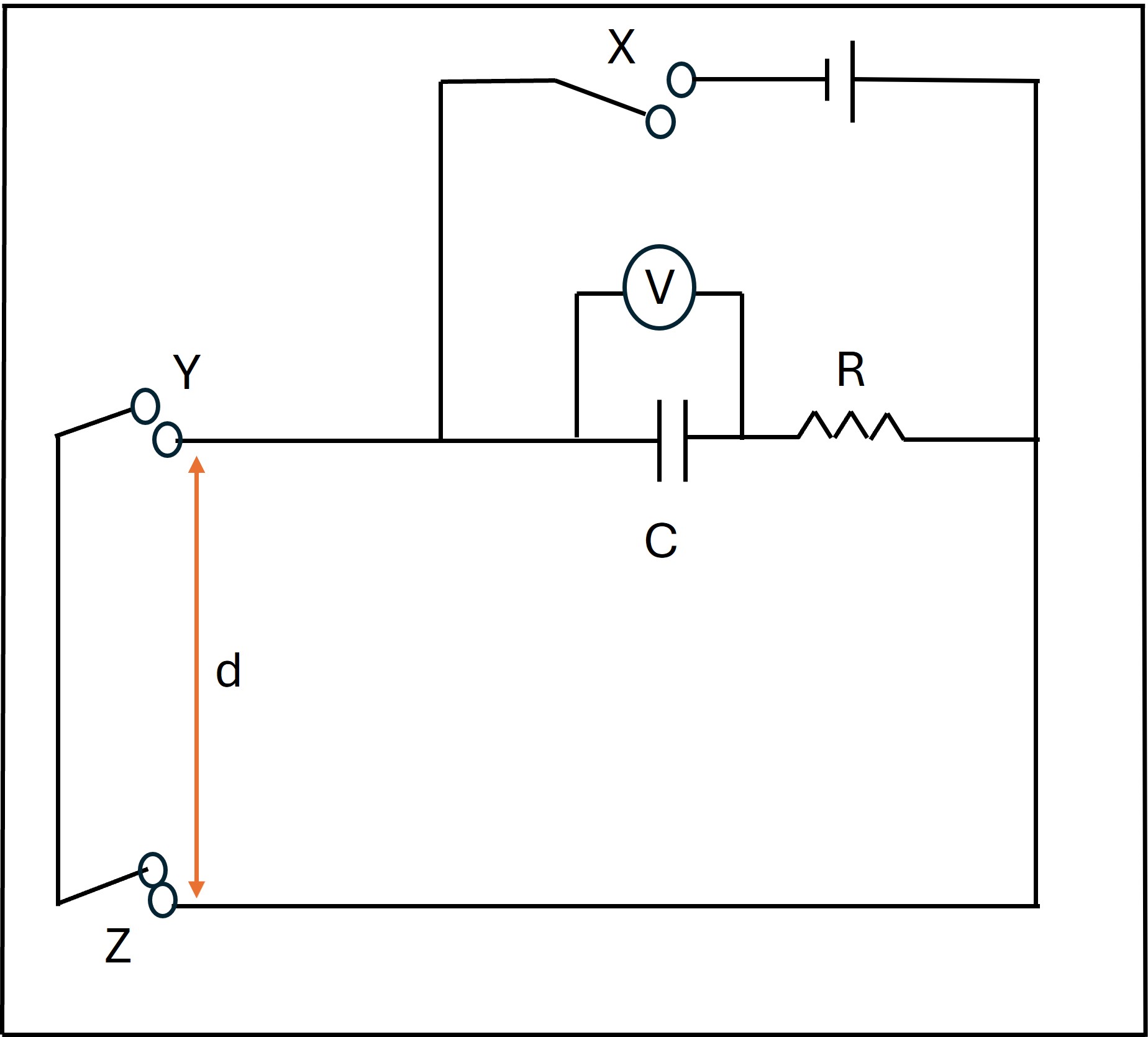}
\label{circuit}
\caption{ The circuit diagram showing the position of the switches $X$, $Y$ and $Z$. The ball undergoes a free fall of distance $d$ between switch $Y$ and $Z$. }
\end{figure}
The working mechanism of switch $Y$  is as follows. The stainless steel ball is put in a holder on a stand. The ball is supported at the bottom by a brass stripe of length 6 cm as shown in figure 2. This stripe is connected to switch $Z$ and can move back and forth in a wooden groove. Another brass stripe of length 2 cm, which is connected to the capacitor, is fixed at one end of the wooden groove. When the 6 cm length stripe is moved quickly sideways in the wooden grove, the ball gets released from the holder and the 6 cm stripe gets in contact with 2 cm stripe as shown in figure 2.  It results in operation of the discharging circuit. Thus, the switch $Y$ turns on when the ball is released. 

\begin{figure}[h]
\centering
\includegraphics[width=6.6cm, height=4.1cm]{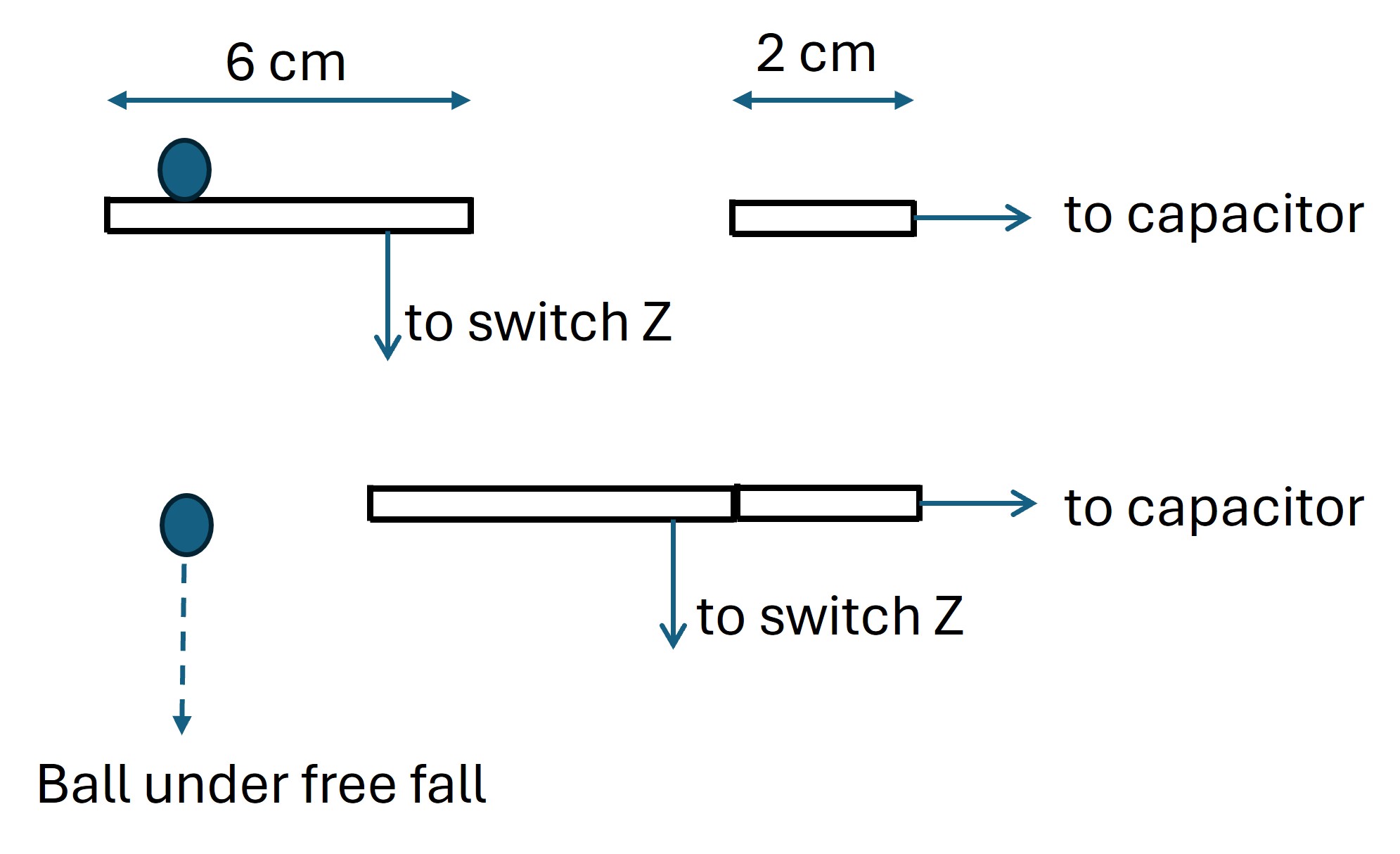}
\label{holder}
\caption{ A schematic diagram showing working of switch $Y$.  }
\end{figure} 
\begin{figure}[h]
\centering
\includegraphics[width=4.8cm, height=4.9cm]{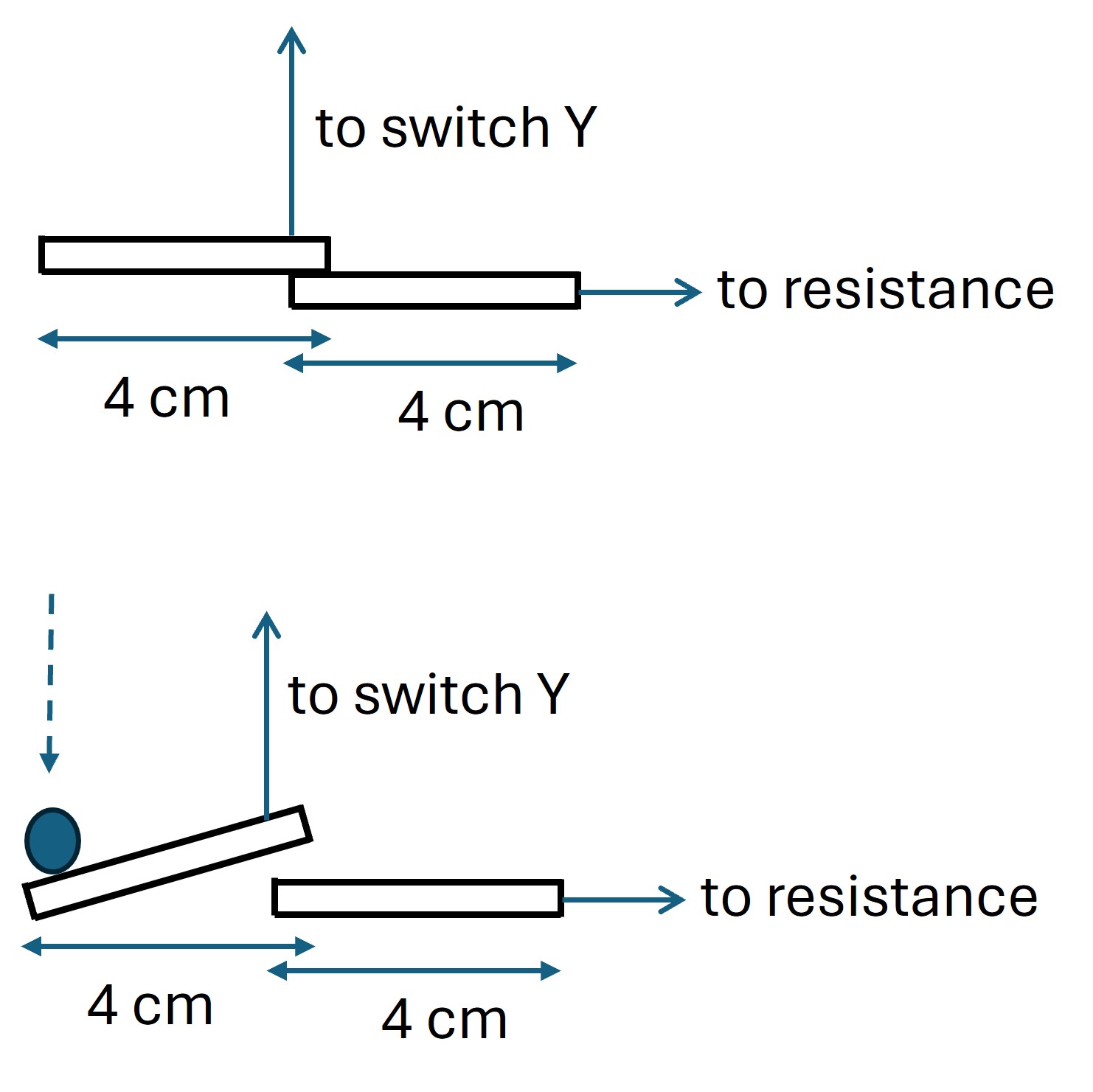}
\label{circuit}
\caption{ A schematic diagram showing working of switch $Z$.  }
\end{figure} 

During the motion of the ball, the voltage of the capacitor decreases. The ball covers distance $d$ under gravity and turns off switch $Z$.  In switch $Z$, a brass tripe of length 4 cm is connected to resistance.  Another brass stripe of 4 cm is connected to switch $Y$ and works similar to a seesaw. These two are in contact as shown in figure 3. When the ball hits brass plate that is connected to switch $Y$, it swings and disconnects the circuit as shown in figure 3.  As a result of it, the discharge of the capacitor halts and the voltage of capacitor reads $V$.  Thus, using eq. (\ref{RC}) and (\ref{g}), g can be calculated.

Since the time of fall in the lab is not more than a meter, the value of resistance and capacitor is chosen such that the time constant is of order of millisecond. Such choice gives considerable drop in voltage for capacitor.

\section{Conclusion}
A low-cost technique to calculate acceleration due to gravity involves a discharging RC circuit. For a free fall of a ball, the time of fall can be calculated by measuring the voltage of capacitor at the beginning and at the end of the fall.   
%%References section
%\section*{References}

\end{document}